\documentclass[aps,prl,floatfix,showpacs,showkeys,twocolumn,superscriptaddress]{revtex4-1}
\newcommand{\sig}{\mbox{\boldmath{$\sigma$}}}
\usepackage{graphicx}
\input{epsf}

\begin{document}

\title{Edge absorption and pure spin current in 2D topological insulator in the Volkov-Pankratov model}

\author{M.M. Mahmoodian}
 \email{mahmood@isp.nsc.ru}
\author{L.I. Magarill}
\author{M.V. Entin}

 \affiliation{Rzhanov Institute of Semiconductor Physics, Siberian Branch of the Russian Academy of Sciences, \\Novosibirsk, 630090, Russia}
 \affiliation{Novosibirsk State University, Novosibirsk, 630090, Russia}

\date{\today}

\begin{abstract}
The light absorption due to the transitions between the edge and two-dimensional (2D) states of a 2D topological insulator (TI) is considered in the Volkov-Pankratov model. It is shown that the transitions are allowed only for the in-plane electric field  orthogonal to the edge of the TI. It is found that the absorption is accompanied by the pure spin photocurrent along the TI edge. The possibility of the  spin current measurement  using polarized luminescence from 2D TI quantum dots is discussed.
\end{abstract}


\maketitle

\subsection{Introduction}

The main purpose of this paper is the study of microwave absorption induced by the transitions from the edge states to 2D states of 2D topological insulator. When the Fermi  energy lies inside the energy gap and the microwave frequency is lower than the 2D absorption threshold only these transitions are responsible for the absorption in a clean sample. Due to the rigid binding between the spin and direction of motion the absorption should be accompanied by the appearance of the edge spin current (details see below).

The edge states in TI were subject of numerous investigations \cite{volk,kane,konig2,zhou,roth,qi,kvon1,shen,kvon2,vova,kvon3,entin,brag}. Most of them rest on the so called Bernevig, Hughes and Zhang (BHZ) Hamiltonian \cite{bern} with zero boundary conditions and its generalizations. In our consideration we will base on the Volkov-Pankratov (VP) \cite{volk} model of the edge states of the 2D topological insulator. The VP model is a minimal model of border states in TI. It rests on the neglect of other energy bands except for the two nearest ones which compose spin- degenerate valence and conduction bands. This determines the value of the model. Unlike the BHZ model, the edge functions in the VP model are non-zero both in ordinary insulator (OI) and in TI.

The VP approach is valid if the distance to the other bands is larger than the OI gap. In a conventional BHZ model it is assumed that the 2D TI is surrounded by the OI (2D or 3D) impenetrable for electrons. The wave function in TI is supposed to satisfy zero boundary condition on the border of TI. This is not the case in the VP model. In a real 2D HgTe the boundary between TI and OI is commonly formed as a result of etching. The TI and OI domains differ by the quantum well (QW) width $d$: $d>d_c$ for TI and $d<d_c$ for OI. In an ordinary 2D TI based on HgTe QW $d_c=6.3$ nm. In our opinion, the OI-TI border is situated inside the QW as a result of smearing of the etching process. As a result, the transition between 2D OI and TI is smooth. This explains why we prefer the VP model.

Besides the absorption we will consider the non-equilibrium pure edge spin current described as a spin flow \cite{bhat,ganich1,ganich2}. This current appears in systems with spin-orbit correlation. Mathematically, this current is expressed via the symmetrized product of operators of spin and velocity averaged with the density matrix.

The pure non-equilibrium spin currents were predicted and observed in a system with multiple GaAs quantum wells \cite{stevens}. After this pioneering observation there were numerous studies of pure spin currents in different systems (see, e.g. \cite{taras,ivch,dyak}). However, till now nobody has studied this current in the edge states of topological insulator where the tie of spin with the direction of motion is ultimate.

\subsection*{Electron states.}

In the 2D case the Hamiltonian of the VP model has the form
\begin{equation}\label{hh}
   H=\left(
       \begin{array}{cc}
         \Delta(y)\sigma_0 &v\sig{\bf k} \\
         v\sig{\bf k} & -\Delta(y)\sigma_0 \\
       \end{array}
     \right),
\end{equation}
where ${\bf k}=(k_x,k_y)$ is the 2D momentum operator, $\sigma_0$ is the $2\times2$ unit matrix, $\sig=(\sigma_x,\sigma_y,\sigma_z)$ are the Pauli matrices.

Below we consider a step-like dependence of $\Delta(y)=\Delta_1 \theta(-y)+\Delta_2 \theta(y)$, $\Delta_1>0$, $\Delta_2<0$, so that the half-plane $y<0$ is an ordinary insulator with  large gap $\Delta_1$, and the half-plane $y>0$ is the TI. In each half-plane $y>0$ and $y<0$ the Hamiltonian (\ref{hh}) has eigenfunctions $\propto e^{i(k_xx+k_yy)}$ with energies
\begin{eqnarray}\label{EcvI}
E=\nu\sqrt{\Delta_{1,2}^2+v^2k^2}.
\end{eqnarray}
Here the index $\nu=\pm1$ corresponds to positive and negative energies; for real ${\bf k}$ the energies get into the conduction and valence bands, respectively. The eigenfunctions are
\begin{eqnarray}\label{wfTI}
&&\Psi^{(1,2)}_{{\bf k},\sigma}=\zeta^{(1,2)}_{{\bf k},\sigma}e^{ik_x x+ik_y y},\\
&&\zeta^{(1,2)}_{{\bf k},+1}=(1,0,0,\alpha^{(1,2)}_{{\bf k},+1}),\\
&&\zeta^{(1,2)}_{{\bf k},-1}=(0,1,\alpha^{(1,2)}_{{\bf k},-1},0),\\
&&\alpha^{(1,2)}_{{\bf k},\sigma}=\frac{v(k_x+i\sigma k_y^{(1,2)})}{E+\Delta_{1,2}}.\label{alpha}
\end{eqnarray}

The wave-functions of the Hamiltonian (\ref{hh}) localized near $y=0$, are composed from the decaying waves with  $k_y^{(1,2)}=-i\lambda_{1,2}$, $\lambda_1>0$ and $\lambda_2<0$ for $y<0$ and $y>0$, correspondingly. Using Eq. (\ref{EcvI}), we find $\lambda_{1,2}^2=(\Delta_{1,2}^2-E^2)/v^2+k_x^2$.

The continuity condition $\Psi^{(1)}_{{\bf k},\sigma}|_{y=0}=\Psi^{(2)}_{{\bf k},\sigma}|_{y=0}$ yields edge states energies $E=\epsilon_{k_x,\sigma}\equiv\sigma vk_x$, parameters $\lambda_{1,2}=\Delta_{1,2}/v$ and the localized edge eigenfunctions
\begin{eqnarray}\label{wfEd}
\psi_{k_x,\sigma}=\frac{C\chi_{k_x,\sigma}}{\sqrt{L_x}}&&e^{ik_x x}\times
\Bigg\{
\begin{array}{ccc}
  e^{\frac{\Delta_1}{v}y}~ & \mbox{at} & y<0, \\
  e^{\frac{\Delta_2}{v}y} & \mbox{at} & y>0,
\end{array}\\
&&\chi_{k_x,+1}=(1,0,0,1),\\
&&\chi_{k_x,-1}=(0,1,-1,0),\\
&&C=\frac12\sqrt{\frac{\Delta_1|\Delta_2|}{v\left(\Delta_1+|\Delta_2|\right)}}.
\end{eqnarray}
These localized solutions are the only ones existing in the domain $|E|<|\Delta_2|$.

If $\Delta_1>|E|>|\Delta_2|$, in addition to the edge states, the 2D solutions appear, delocalized at $y>0$. Their wave-functions should decay at $y\to-\infty$ and combine from two propagating waves at $y>0$ numerated by the continuous momentum $k_y>0$:
\begin{eqnarray}\label{wf2d}
\Psi_{{\bf k},\sigma}=\frac{e^{ik_x x}}{\sqrt{L_x}}\times
\Bigg\{
\begin{array}{ccc}
  C^{(1)}_{{\bf k},\sigma}\zeta^{(1)}_{{\bf k},\sigma}e^{\lambda y}~~~~~~~~~ & \mbox{at} & y<0, \\
  C^{(2)}_{{\bf k},\sigma}\zeta^{(2)}_{{\bf k},\sigma}e^{ik_y y}+\text{c.c.} & \mbox{at} & y>0.
\end{array}
\end{eqnarray}
Here $\lambda=\sqrt{(\Delta_1^2-\Delta_2^2)/v^2-k_y^2}$. The continuity of the wave function at the interface gives
\begin{eqnarray}\label{wf2dcont}
C^{(2)}_{{\bf k},\sigma}=b_{{\bf k},\sigma}C^{(1)}_{{\bf k},\sigma},\\
b_{{\bf k},\sigma}=\frac{\alpha^{(2)*}_{{\bf k},\sigma}-\alpha^{(1)}_{{\bf k},\sigma}}
{\alpha^{(2)*}_{{\bf k},\sigma}-\alpha^{(2)}_{{\bf k},\sigma}},
\end{eqnarray}
where $\alpha^{(1,2)}_{{\bf k},\sigma}$ is defined by Eq.~({\ref{alpha}}) with $E=\nu\varepsilon_k\equiv\nu\sqrt{\Delta_2^2+v^2k^2}$.

Normalizing the wave-function (\ref{wf2d}) we get
\begin{eqnarray}\label{wf2dnorm}
C^{(1)}_{{\bf k},\sigma}=\left[2L_y|b_{{\bf k},\sigma}|^2\left(1+|\alpha^{(2)}_{{\bf k},\sigma}|^2\right)\right]^{-1/2}.
\end{eqnarray}

\begin{figure}[ht]
\leavevmode\centering{\epsfxsize=6.5cm\epsfbox{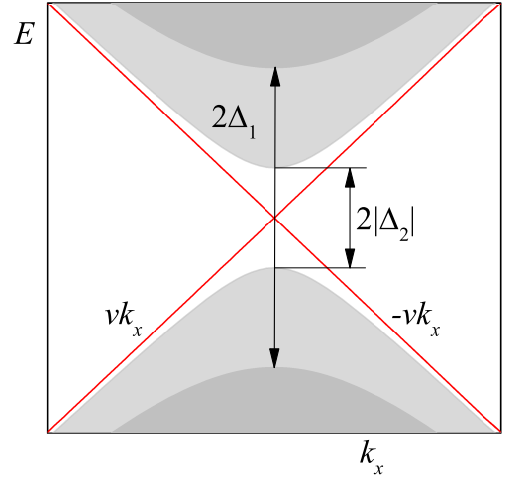}}
\caption{(Color online). Edge spectrum of TI in the step-like VP model for $|\Delta_2|<\Delta_1$. The domains of 2D states of TI and ordinary insulator are filled. At $k_x\to\pm \infty$, edge branches $\pm vk_x$ asymptotically approach the 2D states domain boundaries $\sqrt{\Delta_{1,2}^2+v^2k_x^2}$ and $-\sqrt{\Delta_{1,2}^2+v^2k_x^2}$.}\label{fig1}
\end{figure}

The total energy spectrum is depicted in Fig.~\ref{fig1}.

\subsection*{Probability of direct optical transitions and light absorption.}

The probability of direct transitions between edge and bulk states is given by the matrix elements of electron velocity operator ${\bf V}=\nabla_{\bf k}H$. With the use of wave-functions
Eqs.~(\ref{wf2d}) and (\ref{wfEd}), we get:
\begin{eqnarray}
&&(V_x)_{\nu,{\bf k},\sigma;k_x,\sigma}=0,\label{vx}\\
&&(V_y)_{\nu,{\bf k},\sigma;k_x,\sigma}=-2iv^2C^{(1)}_{{\bf k},\sigma}C\times\nonumber\\
&&~~~~~\times\frac{\left(\Delta_1+|\Delta_2|\right)\left(\nu\varepsilon_k+\epsilon_{k_x,\sigma}+\Delta_1+v\lambda\right)}
{\left(\epsilon_{k_x,\sigma}+v\lambda\right)\left(\nu\varepsilon_k+\epsilon_{k_x,\sigma}\right)\left(\Delta_1+v\lambda\right)}.
\end{eqnarray}

According to Eq.~(\ref{vx}) the $x$-component of the electric field is not absorbed. The probabilities of optical transitions induced by the classical alternating electric field ${\bf E}\cos{(\omega t)}$,  ${\bf E}=(0,E_y,0)$, are given by
\begin{eqnarray}\label{prob}
W_{\nu,{\bf k},\sigma;k_x,\sigma}=\frac{\pi e^2}{2\omega^2}\delta\left(\varepsilon_k-\nu\epsilon_{k_x,\sigma}-\omega\right)\times\nonumber\\
\times|(V_y)_{\nu,{\bf k},\sigma;k_x,\sigma}E_y|^2.
\end{eqnarray}
The light absorption power per unit edge length $Q$ is expressed via the transition probabilities  Eq.~(\ref{prob}):
\begin{eqnarray}\label{abs}
Q=\frac{\omega}{L_x}&&\sum\limits_{\nu,{\bf k},\sigma}G^{(\nu)}_{{\bf k},\sigma}\equiv\omega\sum\limits_\nu g_\nu,
\end{eqnarray}
where the generation temps in the transitions from the edge states to the conduction band $G^{(+1)}_{{\bf k},\sigma}$ and from the valence band to the edge states
$G^{(-1)}_{{\bf k},\sigma}$ are
\begin{eqnarray}
&&G^{(+1)}_{{\bf k},\sigma}=
W_{+1,{\bf k},\sigma;k_x,\sigma}f\left(\epsilon_{k_x,\sigma}\right)\left(1-f\left(\varepsilon_k\right)\right),\label{gtempin}\\
&&G^{(-1)}_{{\bf k},\sigma}=
W_{k_x,\sigma;-1,{\bf k},\sigma}f\left(-\varepsilon_k\right)\left(1-f\left(\epsilon_{k_x,\sigma}\right)\right).\label{gtempout}
\end{eqnarray}
Here $f(E)$ is the Fermi distribution function. At low temperatures and $|E_F|<|\Delta_2|$ ($E_F$ is the Fermi level) we have
\begin{eqnarray}\label{gtemp}
&&G^{(\nu)}_{{\bf k},\sigma}=
W_{\nu,{\bf k},\sigma;k_x,\sigma}\theta\left(\nu\left(E_F-\epsilon_{k_x,\sigma}\right)\right).
\end{eqnarray}
Substituting $G^{(\nu)}_{{\bf k},\sigma}$ from Eq.~(\ref{gtemp}) into Eq.~(\ref{abs}), after integration over ${\bf k}$ we get
\begin{eqnarray}\label{f1}
&&g_\nu=g_0\frac{\delta^2}{\overline{\omega}^2}\Bigg[\frac{\eta_\nu\left(\eta_\nu^2-1\right)\left(\delta-\sqrt{\delta^2-\eta_\nu^2-1}\right)}{\delta\left(\eta_\nu^2+1\right)^2}+\nonumber\\
&&+\arctan\eta_\nu-\left(1-\frac{1}{\delta^2}\right)^2\arctan\left(\frac{\eta_\nu\delta}{\sqrt{\delta^2-\eta_\nu^2-1}}\right)-\\
&&-\frac{\eta_\nu\sqrt{\delta^2-\eta_\nu^2-1}}{\delta^3\left(\eta_\nu^2+1\right)}\Bigg]\theta\left(\overline{\omega}-\overline{\omega}_\nu\right),\nonumber
\end{eqnarray}
where $g_0=e^2E_y^2v/(32\pi\Delta_2^2)$, $\delta=\Delta_1/|\Delta_2|$, $\overline{\omega}=\omega/|\Delta_2|$, $\mu=E_F/|\Delta_2|$, $\overline{\omega}_\nu=\sqrt{\mu^2+1}-\nu\mu$, $\eta_\nu=\sqrt{\overline{\omega}^2-2\nu\mu-1}$.

\noindent In the case of trivial insulator with an infinite gap $\delta\to\infty$ we have
\begin{eqnarray}\label{f2}
g_\nu\approx g_0\frac{2}{\overline{\omega}^2}\left(\arctan\eta_\nu-\frac{\eta_\nu}{\eta_\nu^2+1}\right)\theta\left(\overline{\omega}-\overline{\omega}_\nu\right).
\end{eqnarray}

\begin{figure}[ht]
\leavevmode\centering{\epsfxsize=6.5cm\epsfbox{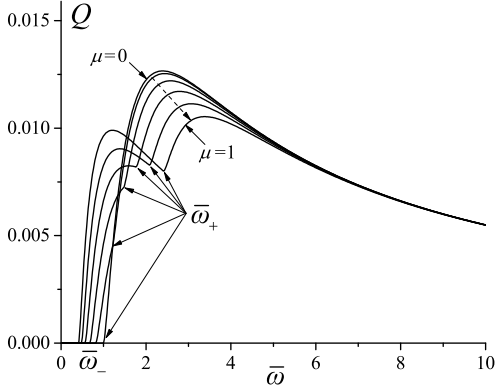}}
\caption{Edge absorption (in units of $e^2E_y^2v/|\Delta_2|$) versus the dimensionless frequency $\overline{\omega}$ for different positions of the dimensionless Fermi level $\mu=0,0.2,0.4,0.6,0.8,1$ (the direction of $\mu$ growth is indicated by dashed arrow). Thin arrows show the upper threshold $\overline{\omega}_+$ positions.}\label{fig2}
\end{figure}

Figure~\ref{fig2} shows the edge absorption as the function of the frequency. Owing to the TI band symmetry, $Q(\mu)=Q(-\mu)$.

\begin{figure}[ht]
\leavevmode\centering{\epsfxsize=6.5cm\epsfbox{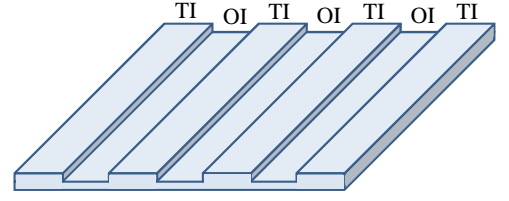}}
\caption{(Color online). Hypothetical structure for measurement of the edge absorption. Wider parts of HgTe layer represent TI, thinner part are OI.}\label{fig3}
\end{figure}

Note, that the absorption exists near the edge of topological insulator only. To observe the edge absorption it is convenient to use an artificial structure with multiple edges, for example with periodically alternating domains of TI and OI, that can be done by alternating quantum well width (See Fig.\ref{fig3}). In such a system the absorbing  power will be proportional to the system area, instead of the edge length.

\subsection*{Spin current}

Owing to the topological protection electron spin number and the direction of motion conserve in the edge states. The phototransitions change the number of edge carriers. At the same time, the numbers of the left and the right-moving carriers are equal. That means the absence of the edge photocurrent in the considered model.

Here we shall consider another effect, namely the spin current. This quantity is determined by the mean product of the operators of spin and electron velocity $(j_s)_{i,j}=\langle\hat{s}_i\hat{v}_j\rangle$. Topological protection makes spin relaxation of edge carriers much longer than that of 2D carriers. Basing on this fact we shall consider the contribution to the edge spin current only. In this approximation the spin current flows within the edge states of TI.

In a specific case of edge electrons and the optical excitation the only component of the spin current exists, $(j_s)_{z,x}$. The velocity operator is reduced to $\sigma v$. The product of the spin number $\sigma$ and velocity, should be averaged with the non-equilibrium part of distribution function of edge  electrons $f_{\sigma,k_x}$. In the absence of backscattering additional numbers of particles excited to or from the edge states are conserved up to carriers reach the contacts. In this ballistic regime all excited carriers contribute the edge spin current. Hence, the ballistic current can be expressed via the total temp of carriers generation in edge states. Partial generation temp from the valence and to conduction bands differs by sign. Formally, we can write a kinetic equation
\[
\sigma v\frac{\partial f_{\sigma,k_x}}{\partial x}=\sum\limits_{\nu,k_y}G^{(\nu)}_{{\bf k},\sigma}.
\]

\begin{figure}[ht]
\leavevmode\centering{\epsfxsize=6.5cm\epsfbox{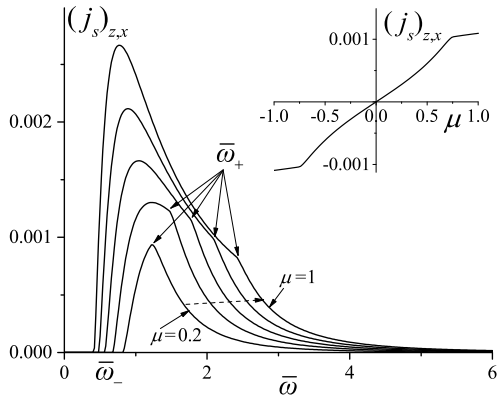}}
\caption{Edge spin current (in units of $e^2E_y^2Lv/\Delta_2^2$) versus the dimensionless frequency $\overline{\omega}$ for different positions of the dimensionless Fermi level $\mu=0.2,0.4,0.6,0.8,1$ (the direction of $\mu$ growth is indicated by dashed arrow). Thin arrows show the upper threshold $\overline{\omega}_+$ positions. Insert: Dependence $j_s(\mu)=-j_s(-\mu)$ at $\overline{\omega}=2$. }\label{fig4}
\end{figure}

Solving the kinetic equation with the conditions $f_{\sigma,p}|_{x=\mp L/2}=0$ on the contacts $\mp L/2$ we get the temp of spin pumping to the contacts:
\begin{eqnarray}\label{jsp}
(j_s)_{z,x}=-\frac{L}{4}\sum\limits_\nu\nu g_\nu,
\end{eqnarray}
where $L$ is the distance between the contacts to the edge. Eq.(\ref{jsp}) shows that the spin current is proportional to the difference of the absorption contributions of transitions from the valence band to the edge state and from the edge states to the conduction band. Fig.~\ref{fig4} demonstrates the dependencies of $(j_s)_{z,x}$ on the frequency and the Fermi level. The band symmetry yields $(j_s)_{z,x}(\mu)=-(j_s)_{z,x}(-\mu)$.

The numerical estimations give the maximal value of $(j_s)_{z,x}=6.5\cdot 10^7$ spins per second. We utilized the quantities $E_y=10$V/cm, $\Delta_2=-20$meV, $L=10^{-3}$cm, $v=5.5\cdot10^7$cm/s.

\begin{figure}[ht]
\leavevmode\centering{\epsfxsize=6.5cm\epsfbox{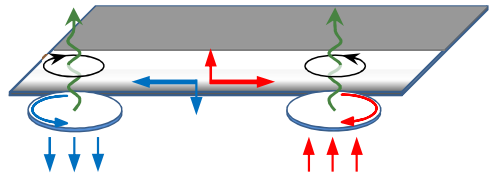}}
\caption{(Color online). A sketch of hypothetic experiment for observation of the edge spin current. The front edge of a TI band is illuminated, the back edge is darkened. The spin current flow along illuminated edge and collected by TI quantum dots where stationary spin polarization occurs. The polarization can be measured, for example, via polarized luminescence.}\label{fig5}
\end{figure}

Let us discuss the hypothetical experiment for the discovery of the edge spin current (see Fig.~\ref{fig5}). To obtain the total edge spin current in TI domain one needs to break the symmetry between the opposite edges. This can be done if to illuminate only one edge remaining the other in darkness. The TI quantum dots connected to this edge will collect spin and produce opposite spin polarization of these dots. The spin polarization can be measured, for example, by the polarized luminescence from these dots. Practically, it would be more suitable to use the same 2D TI material for a large domain and spin-collecting quantum dots. The topological protection yields long-living spin polarization both in large domain and in TI dots. The small number of carriers in the edge states helps to reduce the number of equilibrium carriers in TI dots in such a way that the number of pumped spins becomes comparable with the number of equilibrium carriers. By a choice of abovementioned parameters of TI,  the dot size $10^{-5}$cm, and the spin relaxation time $10^{-6}$s we hope to achieve the spin polarization of dots up to 0.5.

\subsection*{Discussion and conclusions}

In conclusion, we have calculated the edge absorption in a strip of 2D topological insulator due to direct electrodipole transitions between the edge  and 2D valence or conduction bands. The absorption is induced by the inplane electric polarization normal to the edge direction. These phototransitions occur at illumination with frequency below the threshold of direct transitions in 2D domain. Such absorption has been indirectly observed in experiments on the TI photoresponses \cite{kvon3}. We have found that the absorption is accompanied by the edge spin current. The possible way to measure the edge spin currents is offered.

\subsection*{Acknowledgements}
This research was supported by RFBR grant No 17-02-00837. Authors are grateful to E.L. Ivchenko and M.M.~Glazov for fruitful discussions.

\end{document}